\documentclass[11pt,a4paper]{article}
\usepackage{jheppub}
\usepackage{amssymb}
\usepackage{subfigure}
\usepackage{graphicx}
\usepackage{dcolumn,color}
\usepackage{bm}
\usepackage{caption}
\usepackage{epstopdf}
\usepackage{subeqnarray}
\usepackage{epsfig}
\usepackage{color,framed}
\usepackage{amsmath}
\usepackage[table]{xcolor}
\usepackage{cases}
\usepackage{mathrsfs}

\title{Singular P\"oschl-Teller II potentials and gravitating kinks}
\author[]{Yuan Zhong\footnote{Corresponding author.}}

\affiliation[a]{MOE Key Laboratory for Nonequilibrium Synthesis and Modulation of Condensed Matter, School of Physics, Xi'an Jiaotong University, Xi'an 710049, China.}
\affiliation[b]{Institute of Theoretical Physics, Xi'an Jiaotong University, Xi'an 710049, China.}
 \emailAdd{zhongy@mail.xjtu.edu.cn}

\abstract{We report a two-dimensional (2D) gravitating kink model, for which both the background field equations and the linear perturbation equation are exactly solvable. The background solution describes a sine-Gordon kink that interpolating between two asymptotic AdS$_2$ spaces, and can be regarded as a 2D thick brane world solution. The linear perturbation equation can be recasted into a Schr\"odinger-like equation with singular P\"oschl-Teller II potentials. There is no tachyonic state in the spectrum, so the solution is stable against the linear perturbations. Besides, there can be $n=0,1,2,\cdots$ bounded vibrational modes around the kink. The number of these vibrational modes varies with model parameters.
}


\begin{document}
\maketitle


\section{Introduction}

Supersymmetric quantum mechanics (SUSY QM) offers us not only algebraic methods for solving quantum mechanical problems, but also some general conclusions on the structure of the spectrum~\cite{CooperKhareSukhatme1995,GangopadhyayaMallowRasinariu2017}. These methods and conclusions are frequently applied to analyze the linear stability of kink solutions of scalar field theories in both flat~\cite{BoyaCasahorran1989,AndradeMarquesMenezes2020,ZhongLiu2014,ZhongGuoFuLiu2018} and curved spaces~\cite{DeWolfeFreedmanGubserKarch2000,Giovannini2001a,ZhongLiu2013}. 

A key concept in SUSY QM is the superpotential $\mathcal{W}(x)$. Given an arbitrary superpotential, one can always define two partner Hamiltonian operators: $H_{\pm}= -\frac{d^2}{dx^2}+V_{\pm}(x)$ with $V_{\pm}(x)=\mathcal{W}^2\pm\frac{d\mathcal{W}}{dx}$. Both of the Hamiltonians can be factorized as products of a linear differential operator $\mathcal{A}\equiv \frac{d}{dx}+\mathcal{W}$ and its hermitian conjugate $\mathcal{A}^\dagger= -\frac{d}{dx}+\mathcal{W}$: 
\begin{equation}
H_{+}=\mathcal{A}\mathcal{A}^\dagger,\quad H_{-}=\mathcal{A}^\dagger\mathcal{A}.\nonumber
\end{equation}
Usually, if $V_{\pm}(x)$ are both smooth, the spectra of $H_{\pm}$ are non-negative, and except the ground state, the structures of the spectra are equivalent~\cite{CooperKhareSukhatme1995,GangopadhyayaMallowRasinariu2017}. 
Further, if the partner potentials  $V_{\pm}(x)$ are shape invariant, i.e., $V_{+}(x ; a_{0})=V_{-}(x ; a_{1})+F(a_{1})$, where $a_0, a_1$ are two sets of constant parameters, and $F(a_1)$ is a function of $a_1$, then  the corresponding Schr\"odinger equations are exactly solvable~\cite{Gendenshtein1983,DuttKhareSukhatme1988,BougieGangopadhyayaMallow2010}. 

The so called P\"oschl-Teller (PT) II  potentials~\cite{PoschlTeller1933} are a pair of partner potentials generated by the following superpotential~\cite{CooperKhareSukhatme1995,GangopadhyayaMallowRasinariu2017}:
\begin{equation}
\mathcal{W}(x; l,m)=l\tanh{x}-m \coth x,\quad l>m,
\end{equation} where $l$ and $m$ are two real parameters.
One can easily check that the partner potentials
\begin{eqnarray}
V_{\textrm{PT},+}(x;l,m)&=&(l-m)^{2}-l(l-1) \operatorname{sech}^{2}(x)+m(m+1) {\rm{csch}}^{2}(x) \\
V_{\textrm{PT},-}(x;l,m)&=&(l-m)^{2}-l(l+1) \operatorname{sech}^{2}(x)+m(m-1) {\rm{csch}}^{2}(x)
\end{eqnarray}
are shape invariant: $V_{\textrm{PT},+}(x; l,m)=V_{\textrm{PT},-}(x; l-1,m+1)+(l-m)^2-(m-l+2)^2$, and therefore, are exactly solvable.

When parameter $m=0$, the PT II potentials  are smooth and finite, as can be seen from their expressions:
\begin{eqnarray}
V_{\textrm{PT},+}(x;l,0)&=&l^{2}-l(l-1) {\rm{sech}}^{2}(x), \\
V_{\textrm{PT},-}(x;l,0)&=&l^{2}-l(l+1) {\rm{sech}}^{2}(x).
\end{eqnarray}  
This regular version of PT II potentials is frequently seen in equations for linear perturbations  around many static solutions. For example, the linear perturbations around the sine-Gordon kink and the $\phi^4$ kink in flat space satisfy Schr\"odinger-like equations with $V_{\textrm{PT},-}(x;1,0)$ and $V_{\textrm{PT},-}(x;2,0)$, respectively~\cite{DashenHasslacherNeveu1974,GoldstoneJackiw1975,Rajaraman1975}. The solvability of these potentials makes it possible to give some quantitative discussions on the quantization~\cite{DashenHasslacherNeveu1974,GoldstoneJackiw1975,Rajaraman1975,Jackiw1977,BoyaCasahorran1989,Evslin2021} and dynamical collision phenomena of kinks~\cite{Sugiyama1979,Moshir1981,CampbellSchonfeldWingate1983,AnninosOliveiraMatzner1991,DoreyMershRomanczukiewiczShnir2011,TakyiWeigel2016,AdamOlesRomanczukiewiczWereszczynski2019a,MantonOlesRomanczukiewiczWereszczynski2021}, see~\cite{Vachaspati2006} for a comprehensive review. 
Perturbation equations of some de Sitter thick branes~\cite{Wang2002,AfonsoBazeiaLosano2006,LiuZhongYang2010} or black holes solutions~\cite{FerrariMashhoon1984,FerrariMashhoon1984a,MossNorman2002,CardosoLemos2003,MolinaGiugnoAbdallaSaa2004,Jing2004,MolinaNeves2012} are also closely related to the regular PT II potentials.

When $m\neq 0$, at least one of the partner potentials will become singular. These singularities may cause an explicit  breaking of supersymmetry between $H_{\pm}$, and make their spectra different, even worse, bound states with negative eigenvalues can appear, see Ref.~\cite{CasahorranNam1991} for discussions on the case with $m=1$. Although singular PT II potentials have been found in the perturbation equations of some black hole solutions~\cite{DuWangSu2004,CardonaMolina2017,BhattacharjeeSarkarBhattacharyya2021}, it is still unclear if such potentials are linked to any kink solution.

In fact, singular potentials have been found in the linear perturbation equations of some gravitating kink solutions, such as those in Refs.~\cite{Stoetzel1995,Zhong2021,ZhongLiLiu2021,FengZhong2022}, where the kinks are embedding in some asymptotic AdS$_2$ spaces, and can be regarded as two-dimensional thick brane world solutions. Thus, it would be a natural conjecture that in some gravitating kink models, the linear perturbations are described by singular PT II potentials. However, constructing such kink models is not easy. Because to write the linear perturbation equation of a gravitating kink into a Schr\"odinger-like equation, one usually needs to introduce at least a metric dependent coordinate transformation~\cite{Zhong2021}. If the scalar field that generates the kink is a K-field, namely, scalar fields with non-minimal first-order derivative couplings~\cite{Armendariz-PiconMukhanovSteinhardt2001,Armendariz-PiconDamourMukhanov1999,GarrigaMukhanov1999}, then another coordinate transformation is needed~\cite{ZhongLiLiu2021,Zhong2021b}. Usually, these coordinate transformations are rather complicated, it is impossible even to obtain the explicit expression of the Schr\"odinger-like equations (see~\cite{Zhong2021,ZhongLiLiu2021,FengZhong2022}), needless to say to solve them exactly.

In this work, we consider a 2D gravitating kink model with a scalar K-field. We show that under certain conditions, the coordinate transformation caused by the derivative coupling of the K-field inverses the one caused by the metric. In this case, we can explicitly write out the  Schr\"odinger-like equation for the linear perturbation without introducing any coordinate transformation. Moreover, we show that for the sine-Gordon kink, the linear perturbations are described by a Schr\"odinger-like equation with $V_{\textrm{PT},+}(x;l,1)$, where the value of parameter $l$ depends on the value of gravitational coupling and the expectation value of the vacuum. By tuning the value of $l$, we would have an arbitrary number of discrete bound states in the spectrum.  To the author's knowledge, this is the first 2D gravitating kink solution that has exactly solvable linear perturbation equation. This solution might be useful for considering the quantization and collisions of gravitating kinks.

This work is organized as follows. In the next section, after a briefly description of our model, we show that the field equations have simple first-order formalism, from which analytical kink solutions can be easily constructed. In section \ref{secLinP}, we first review  the linear stability issue of 2D gravitating kink. Then, we derive the condition under which the linear perturbation equation can be explicitly expressed without introducing coordinate transformations. Then, we move to an exactly solvable case in section \ref{secSolution}. We end our work with some concluding remarks.

\section{The model and the first-order equations}
\label{SecTwo}
Following~\cite{ZhongLiLiu2021}, we consider a 2D gravitating K-field model with the following action:
\begin{equation}\label{1}
S=\frac{1}{\kappa} \int d^{2} x \sqrt{-g}\left[-\frac{1}{2} \nabla^{\mu} \varphi \nabla_{\mu} \varphi+\varphi R+\kappa \mathcal{L}(\phi,X)\right],
\end{equation}
where  $\kappa>0$ is the gravitational coupling constant, $\varphi$ is the dilaton field which couples with the curvature scalar $R$, and $X\equiv-\frac12\nabla^\mu\phi\nabla_\mu\phi$ is the kinetic term of the scalar matter field $\phi$. Note that the Lagrangian density of a K-field can always be written as a function of $\phi$ and $X$.

We will looking for static gravitating kink solutions with the following  metric:
\begin{equation}
\label{metricXCord}
  ds^2=-e^{2A(x)}dt^2+dx^2,
\end{equation}
where $A(x)$ is called the warp factor~\cite{RandallSundrum1999,RandallSundrum1999a}. With this metric, the dilaton equation reduces to a simple algebraic equation~\cite{Stoetzel1995,Zhong2021,ZhongLiLiu2021,FengZhong2022}:
\begin{equation}
\label{dilatonEq}
\varphi(x)=2A(x),
\end{equation}
and the independent dynamical equations we need to solve are the Einstein equations:
\begin{eqnarray}
\label{eqEin1}
-4 \partial_x^2 A&=&\kappa  \mathcal{L}_X (\partial_x\phi)^2, \\
\label{eqEin2}
4 \partial_x^2 A +2 \left(\partial_x A\right)^2&=&\kappa  \mathcal{L},
\end{eqnarray}
where $\mathcal{L}_{X}\equiv \frac{\partial\mathcal{L}}{ \partial{X}}$. Depending on the form of $\mathcal{L}(\phi,X)$, there can be more unknown variables than independent equations, in this case, we need to impose some constraints. 

For the canonical case $\mathcal{L}=\bar{\mathcal{L}}\equiv X-V(\phi)$, there are three unknown variables of the system: $A(x)$, $\phi(x)$ and $V(\phi)$. Thus, we have to impose a constraint equation. It  is  convenient to introduce a function $W(\phi)$, whose form can be chosen arbitrarily, and constrain the kinetic term of the scalar field via
\begin{eqnarray}
\label{FirCan1}
X=-\frac{1}{2}W_\phi^2,
\end{eqnarray} or equivalently,
\begin{eqnarray}
\label{FirCan1b}
\partial_x\phi= W_\phi\equiv\frac{d W(\phi)}{d \phi}.
\end{eqnarray} 
Under this constraint, eqs. \eqref{eqEin1} and \eqref{eqEin2} give~\cite{Stoetzel1995,Zhong2021,FengZhong2022}
\begin{eqnarray}
\label{FirCan2}
\partial_x A&=&-\frac14\kappa W,\\
\label{FirCan3}
V&=&\frac{1}{2} W_\phi^2-\frac{1}{8} \kappa  W^2.
\end{eqnarray} 
If $W(\phi)$ is properly chosen, one can derive exact gravitating kink solutions from the first-order equations~\eqref{FirCan1b}-\eqref{FirCan3}, see refs.~\cite{Stoetzel1995,Zhong2021,FengZhong2022,LimaAlmeida2022} for examples. Note that eq.~\eqref{FirCan2} also indicates the following relation
\begin{equation}
\label{EqAPhi}
A(\phi)=-\frac14\kappa\int \frac{W}{W_\phi}d\phi,
\end{equation}
which will be used later.

For noncanonical scalars, the Einstein equations usually do not have simple first-order formalism, an exception can be found in~\cite{ZhongLiLiu2021}. In present work, we consider a K-field model with the following Lagrangian density~\cite{AdamQueiruga2011,ZhongLiu2015}:
\begin{equation}
\label{TwinAction}
\mathcal{L}=U(\phi)(X+\frac12W_\phi^2)^2+X-V(\phi),
\end{equation}
where $W(\phi)$ is the same function introduced in eq.~\eqref{FirCan1}. This model has one more unknown variable than the canonical model, i.e.,  $U(\phi)$. Thus, we need to impose two constraints.  If we also constrain $X=-\frac{1}{2}W_\phi^2$, as in the canonical case, then the Einstein equations~\eqref{eqEin1}-\eqref{eqEin2} take the same form as those of the canonical model, because
\begin{equation}
\label{twinCond1}
\mathcal{L}_ X\big\vert_{X=-\frac{1}{2}W_\phi^2}=1=\bar{\mathcal{L}}_X,
\end{equation}
and 
\begin{equation}
\label{twinCond2}
\left.\mathcal{L}\right\vert_{X=-\frac{1}{2}W_\phi^2}=X-V(\phi)=\bar{\mathcal{L}}.
\end{equation}
Consequently, the first-order equations~\eqref{FirCan1b}-\eqref{FirCan3} can also be used to construct solutions of the K-field model of \eqref{TwinAction}. We call \eqref{TwinAction} a twinlike model of the canonical model, see~\cite{AndrewsLewandowskiTroddenWesley2010,AdamQueiruga2011,BazeiaDantasGomesLosanoEtAl2011,BazeiaMenezes2011,AdamQueiruga2012,BazeiaDantas2012,BazeiaLobMenezes2012,BazeiaLobaoLosanoMenezes2014,ZhongLiu2015,ZhongFuLiu2018,AdamVarela2020,DantasRodrigues2020,BazeiaLosanoMarquesMenezes2020} for details and more examples of twinlike models. 

The results obtained so far are irrelevant of $U(\phi)$, which will become important as soon as the linear perturbation issue is considered.

\section{Linear perturbations}
\label{secLinP}
The linear perturbation issue for solutions of action \eqref{1} with metric \eqref{metricXCord} has been discussed in~\cite{ZhongLiLiu2021,Zhong2021b}. In this section, we first review the main steps of~\cite{ZhongLiLiu2021}, and then apply the results to the model of~\eqref{TwinAction}.

 To start with, we introduce the conformal flat coordinates:
\begin{equation}
  ds^2=e^{2A(r)}(-dt^2+dr^2),
\end{equation}
where
\begin{equation}
r\equiv\int e^{-A(x)}dx.
\end{equation}
For simplicity, we will use overdots and primes to denote the derivatives with respect to $t$ and $r$, respectively. 
Suppose that $\{\varphi(r), \phi(r), g_{\mu\nu}(r)\}$ describe an arbitrary static solution of the field equations, and $\{\delta\varphi(r,t), \delta\phi(r,t), \delta g_{\mu\nu}(r,t)\}$ are small perturbations around the solution, where
\begin{eqnarray}
\delta g_{\mu\nu}(r,t)&\equiv& e^{2A(r)} h_{\mu\nu}(r,t)\nonumber\\
&=&e^{2A(r)} \left(
\begin{array}{cc}
 h_{00}(r,t) & \Phi (r,t) \\
 \Phi (r,t) & h_{rr}(r,t) \\
\end{array}
\right).
\end{eqnarray}
To fix the  gauge degrees of freedom in the linear perturbation equations, we define a new variable $\Xi \equiv 2 \dot{\Phi}-h_{00}^{\prime}$, and take $\delta \varphi=0$ as the gauge condition~\cite{ZhongLiLiu2021}.

The linearization of the field equations leads to three independent equations. Two of them enable us to eliminate $\Xi$ and $h_{rr}$ in terms of $\delta\phi$, and the other, after eliminating $\Xi$ and $h_{rr}$, takes the following form~\cite{ZhongLiLiu2021}:
\begin{eqnarray}
&&-\ddot{\delta \phi }
+\gamma  \delta \phi ''
+\gamma  \left(\frac{\gamma '}{\gamma}
+\frac{  \mathcal{L}_{X}'}{{\mathcal{L}_{X}}}\right)\delta \phi ' +\gamma
	 \left[
	 	A'^2
		-3 A''
		+2\frac{ A'''}{A'}
		- 2 \left(\frac{A''}{A'}\right)^2\right.\nonumber\\
		&&\left.
		-\frac12 \frac{\mathcal{L}_X ' }{ \mathcal{L}_X } \frac{X'}{ X}
		-\frac12\frac{X''}{ X}
		+\frac{1}{4} \left(\frac{X'}{X}\right)^2
	\right]\delta \phi +\gamma '
  \left(
	\frac{A''}{A'}
	-\frac12\frac{X'}{ X}	-A' \right)\delta \phi =0,
\end{eqnarray}
where
\begin{equation}
\gamma\equiv 1+2\frac{\mathcal{L}_{XX}X}{\mathcal{L}_{X}},\quad \mathcal{L}_{XX}\equiv\frac{\partial^2\mathcal{L} }{\partial X^2}.
\end{equation}

By rescaling the variable
\begin{equation}
\label{rescaling}
G(r,t) \equiv  \mathcal{L}_X^ {1/2}\gamma^{1/4 }  \delta\phi(r,t),
\end{equation}
and introducing another coordinate transformation $r\to y= \int dr \gamma^{-1/2}$, we obtain a compact wave equation~\cite{ZhongLiLiu2021}:
\begin{equation}
\label{eqWave}
-\partial_{t}^2 G+\partial_{y}^2 G- V_{\text{eff}}(y)G =0,
\end{equation}
where
\begin{equation}
V_{\text{eff}}(y)\equiv \frac{\partial_y^2 f}{f},
\end{equation}
and
\begin{equation}
 f(y)\equiv \mathcal{L}_X^{1/2} \gamma ^{1/4} \frac{\partial_y\phi}{\partial_y A}.
\end{equation}
 Note that in order the rescaling \eqref{rescaling} and the coordinate transformation $r\to y$ are meaningful, the scalar Lagrangian must satisfies the following conditions:
\begin{equation}
\label{stabilityCond}
\mathcal{L}_X>0,\quad \gamma>0.
 \end{equation}
 After conducting the modes expansion
\begin{equation}
\label{eqModeExpan}
G(y,t)=\sum_n e^{i\omega_n t}\psi_n(y),
\end{equation}
eq.~\eqref{eqWave} becomes a Schr\"odinger-like equation of $\psi_n(y)$~\cite{ZhongLiLiu2021}:
\begin{equation}
\label{eqSch}
{H} \psi_n\equiv -\frac{d^2\psi_n}{dy^2}+V_{\text{eff}}(y)\psi_n=\omega_n^2 \psi_n.
\end{equation}
The above Hamiltonian operator can be factorized as follows~\cite{ZhongLiLiu2021}:
\begin{equation}
H ={ \mathcal{A}}{ \mathcal{A}}^\dagger,
\end{equation}
where
\begin{equation}
{\mathcal{A}}=\frac{d}{d y}+\frac{\partial_y f}{f}, \quad {\mathcal{A}}^{\dagger}=-\frac{d}{d y}+\frac{\partial_y f}{f}.
 \end{equation}
 We see that the superpotential is $\mathcal{W}(x)=\frac{\partial_y f}{f}$.

As we said in the introduction, the factorization of a Hamiltonian operator usually ensures the semipositive definite of  the eigenvalues, i.e., $\omega_n^2\geq 0$. In this case, we say the background kink solution is stable against the linear perturbations.
But if $V_{\text{eff}}(y)$ is singular, the factorization of a Hamiltonian operator dose not necessarily indicate the stability of the solution. The meaning of this statement will become clear, if we can find a gravitating kink solution with exactly solvable linear perturbation equation. The problem is that even if we can easily construct exact kink solutions in the $x$-coordinates by using the first-order equations~\eqref{FirCan1b}-\eqref{FirCan3}, the complicated coordinate transformations from $x\to r$ and $r\to y$  would make it difficult, in general, to obtain the explicit expression of the Schr\"odinger-like equation~\eqref{eqSch}, needless to say to solve it.

We noticed that it is possible to construct K-field models, such that the transformation from $r\to y$ inverses the one from $x\to r$. This is a very interesting situation, where both the transformations from $x\to r$ and from $r\to y$ are non-trivial, but their combination is a trivial identity transformation:
\begin{equation}
\label{CTcondition}
\frac{dy}{dx}= e^{-A}\gamma^{-1/2}=1.
\end{equation}
In this case, the Schr\"odinger-like equation~\eqref{eqSch} can be obtained directly in the $x$-coordinates:
\begin{equation}
{H} \psi_n\equiv -\frac{d^2\psi_n}{dx^2}+V_{\text{eff}}(x)\psi_n=\omega_n^2 \psi_n,
\end{equation}
with
\begin{equation}
V_{\text{eff}}(x)\equiv \frac{\partial_x^2 f}{f},
\end{equation}
and
\begin{equation}
\label{xGroundWave}
 f(x)\equiv \mathcal{L}_X^{1/2} \gamma ^{1/4} \frac{\partial_x\phi}{\partial_x A}.
\end{equation}

In fact, the K-field model in~\eqref{TwinAction} can realize the above idea, if $U(\phi)$ is properly chosen. For this model
\begin{equation}
\label{eqGammaUW}
\gamma\vert_{X=-\frac{1}{2}W_\phi^2}=1-2U(\phi)W_\phi^2,
\end{equation}
so the condition in eq.~\eqref{CTcondition} can be satisfied if 
\begin{equation}
\label{eqGammaA}
\gamma\vert_{X=-\frac{1}{2}W_\phi^2}=e^{-2A},
\end{equation}
in other words,
\begin{equation}
\label{eqU}
U(\phi)=\frac{1}{2W_\phi^2}(1-e^{\frac\kappa 2\int \frac{W}{W_\phi}d\phi}).
\end{equation}
To obtain the last equation we used eq.~\eqref{EqAPhi}. 

The discussion so far is valid for arbitrary function $W(\phi)$. After specify the form of $W(\phi)$, the Lagrangian in \eqref{TwinAction} will be completely fixed, as $U(\phi)$ and $V(\phi)$ can be calculated from eq.~\eqref{eqU} and eq.~\eqref{FirCan3}, respectively. The solutions of $\phi(x)$ and $A(x)$ will be obtained by solving  eqs.~\eqref{FirCan1b}-\eqref{FirCan2}.
Then, by substituting the background solution into eq.~\eqref{xGroundWave} and using eqs.~\eqref{twinCond1} and \eqref{eqGammaA} we obtain
\begin{equation}
\label{eqfSimple}
 f\vert_{X=-\frac{1}{2}W_\phi^2}=e^{-A/2} \frac{\partial_x\phi}{\partial_x A},
\end{equation}
with which the effective potential $V_{\text{eff}}(x)\equiv \frac{\partial_x^2 f}{f}$ of the Schr\"odinger-like equation can be calculated.  Now, let us turn to an explicit example.

\section{Sine-Gordon kink and the PT II potentials}
\label{secSolution}

The sine-Gordon kink is a typical kink solution, which can be obtained by taking
\begin{equation}
W(\phi)=k v^2 \sin\left(\frac{\phi}{v}\right),
\end{equation}
for which $U(\phi)$ and $V(\phi)$ take the following form:
\begin{eqnarray}
U(\phi)&=&-\frac{1}{2 k^2 v^2}\sec ^2\left(\frac{\phi }{v}\right) \left[\cos ^{-\frac{1}{2} \kappa  v^2}\left(\frac{\phi }{v}\right)-1\right],\\
V(\phi)&=&\frac{1}{16} k^2 v^2 \left[\left(\kappa  v^2+4\right) \cos \left(\frac{2 \phi }{v}\right)-\kappa  v^2+4\right].
\end{eqnarray}
The solution of eqs.~\eqref{FirCan1b}-\eqref{FirCan2} read
\begin{eqnarray}
\label{SolPhi}
\phi(x)&=&v \arcsin(\tanh(k x)),\\
\label{SolWarp}
A(x)&=&-\frac{1}{4} \kappa  v^2 \log (\cosh (k x)).
\end{eqnarray}
Obviously, the parameters $k$ and $v$ describe the thickness of the kink and the expectation value of the vacua, respectively. This solution can be regarded as a 2D thick brane solution, as the metric is asymptotically AdS$_2$, see~\cite{DzhunushalievFolomeevMinamitsuji2010} for comprehensive of thick brane models in 5D.

\begin{figure*}[!ht]
\centering
\includegraphics[width=1\textwidth]{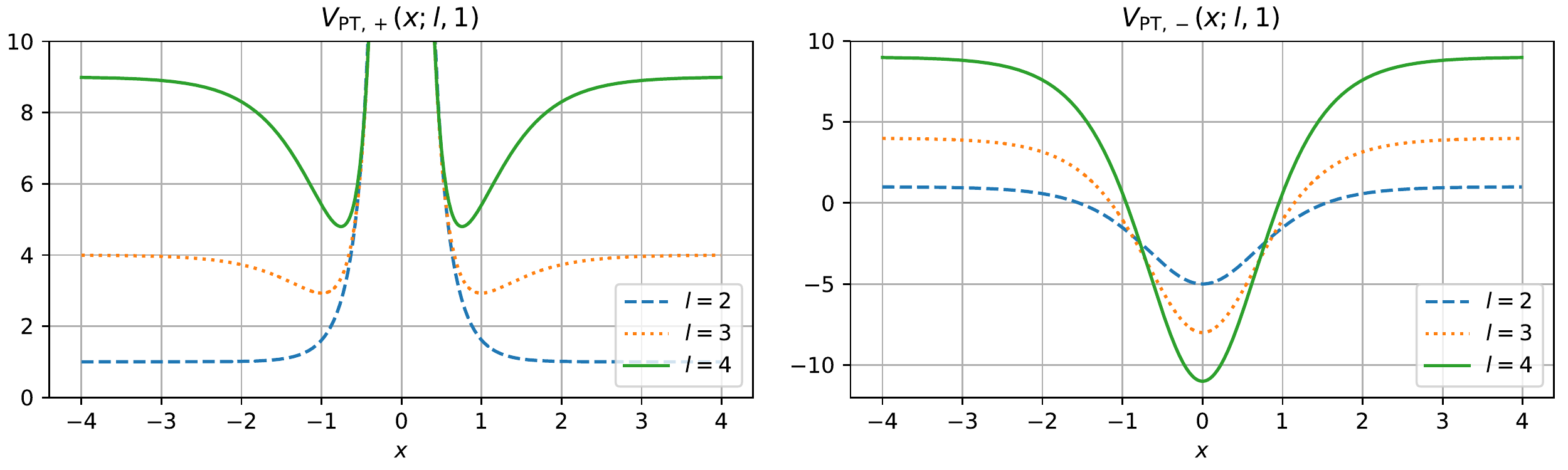}
\caption{Plots of the effective potential $V_{\rm{eff}}(x)=V_{\textrm{PT},+}(x;l,1)$ and its partner potential $V_{\textrm{PT},-}(x;l,1)$ for $l=2,3,4$. }
\label{Fig1}
\end{figure*}

From eq.~\eqref{eqfSimple} we get
\begin{equation}
f(x)=-\frac{4 \text{csch}(k x) \cosh ^{\frac{\kappa  v^2}{2}}(k x)}{\kappa  v}.
\end{equation}
After denoting $l\equiv \frac{\kappa  v^2}{2}>0$, and taking $k=1$, we obtain the expression of the effective potential:
\begin{equation}
\label{PTpotential1}
V_{\rm{eff}}(x)\equiv \frac{\partial_x^2 f}{f}=(l-1)^2-l (l-1) \text{sech}^2(x)+2 \text{csch}^2(x),
\end{equation}
which is nothing but one of the P\"oschl-Teller II potential $V_{\textrm{PT},+}(x;l,m)$ with $m=1$.
The corresponding partner potential is
\begin{equation}
\label{PTpotential2}
V_{\textrm{PT},-}(x;l,1)= (l-1)^2- l (l+1) \text{sech}^2( x).
\end{equation}
In figure~\ref{Fig1}, we plotted the shapes of $V_{\textrm{PT},+}$ and $V_{\textrm{PT},-}$ for $m=1$ and $l=2,3,4.$ We see that $V_{\textrm{PT},+}(x;l,1)$ is always divergent at $x=0$, while $V_{\textrm{PT},-}(x;l,1)$ is smooth. Besides, $V_{\textrm{PT},+}(x;l,1)$ is positive everywhere, which means the background kink solution is stable for these parameters.

The properties of these two potentials have been discussed in ref.~\cite{CasahorranNam1991}.
For both potentials, the continuous spectrum starts from $(l-1)^2$. When the parameter $l$ is large enough, there can be some bound states for both potentials. The spectra of the bound states for $V_{\textrm{PT},+}(x;l,1)$ and $V_{\textrm{PT},-}(x;l,1)$ are~\cite{CasahorranNam1991}
  \begin{equation}
E_{+,n}^{(l,1)}= (l-1)^2-(l-2n-3)^2,\quad n=0,1,2,\cdots<\frac{l-3}{2},
\end{equation}
and 
  \begin{equation}
E_{-,n}^{(l,1)}=(l-1)^2-(l-n)^2,\quad n=0,1,2,\cdots<l, \quad l>1,
\end{equation}
respectively. Here, instead of $\omega^2$, we used $E$ to represent the eigenvalues, just for simplicity. Obviously, the degeneracy between the spectra are only partly reserved: $E_{+,n}^{(l,1)}=E_{-,2n+3}^{(l,1)}$.

When $l>3$, $V_{\textrm{PT},+}(x;l,1)$ can capture bound states. For $l$ even or odd, $l=2p$ or $l=2p+1$ for some positive integer $p$, the number of bound states is $p-1$~\cite{CasahorranNam1991}. All the bound states of $V_{\textrm{PT},+}(x;l,1)$, when existing, have positive eigenvalues. Especially, the ground state has an eigenvalue $E_{+,0}^{(l,1)}= 4l-8$. On the other hand, $V_{\textrm{PT},-}(x;l,1)$ always have a bound state with negative eigenvalue  $E_{-,0}^{(l,1)}=1 - 2 l$, and a zero mode $E_{-,1}^{(l,1)}=0$.

\begin{figure*}[!ht]
\centering
\includegraphics[width=0.7\textwidth]{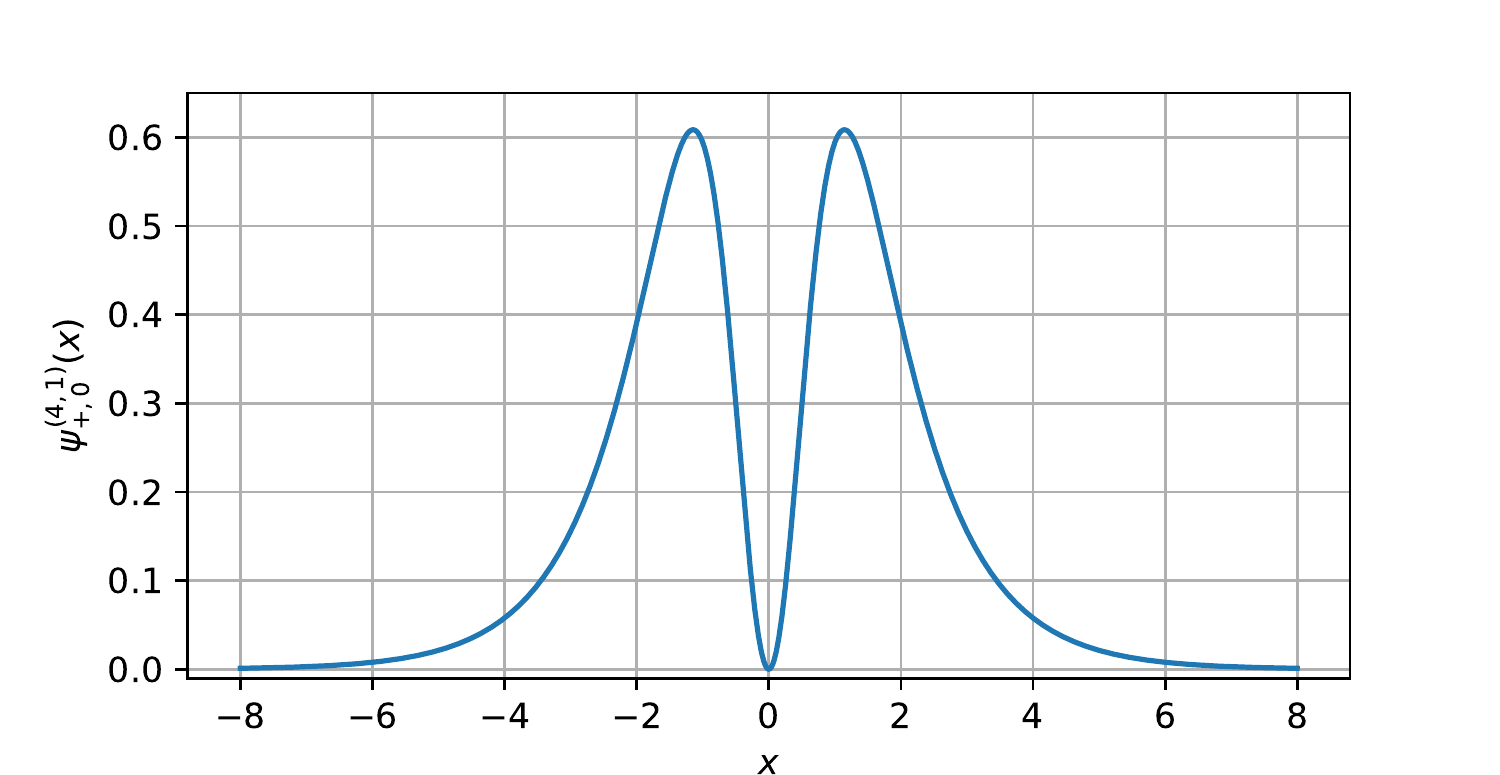}
\caption{Wave function of the bound state of $V_{\rm{eff}}(x)=V_{\textrm{PT},+}(x;4,1)$. }
\label{Fig2}
\end{figure*}

For example, when $l=4$, $V_{\textrm{PT},-}(x;4,1)$ has four bound states, the corresponding eigenvalues are $E_{-,n}^{(4,1)}=-7, 0, 5, 8$, for $n=0,1,2,3$, respectively. While $V_{\textrm{PT},+}(x;4,1)$ only have one bound state with eigenvalue $E_{+,0}^{(4,1)}=8$. The corresponding wave function is
 \begin{equation}
  \psi_{+,0}^{(4,1)}(x)=\sqrt{\frac{5}{2}} \tanh ^2(x) \text{sech}(x).
  \end{equation}
From eq.~\eqref{eqModeExpan} we know that this bound state corresponds to a vibrational modes, with frequency $\omega_0=2\sqrt{2}$, around the kink. Note that the wave function $\psi_{+,0}^{(4,1)}(x)$ is vanished at $x=0$, see figure~\ref{Fig2}. This behavior of wave function is caused by the hard singularity of the effective potential. Here, by hard singularity, we mean a term $\sim \frac{\alpha}{x^2}$ with $\alpha\geq \frac34$~\cite{FrankLandSpector1971}. One may check that when $x\to0$ the singular PT II potentials $V_{\textrm{PT},+}(x;l,1)$ scaling as $\frac{2}{x^2}$. Therefore, $V_{\textrm{PT},+}(x;l,1)$ has a hard singularity at $x=0$, which will force $\psi_{+,n}^{l,1}(0)=0$, and the range of $x$ is effectively broken into two disjoint pieces $x < 0$ and $x > 0$ with no communication between them~\cite{CooperKhareSukhatme1995}.

\section{Concluding remarks}
In this work, we studied a special K-field model in a two-dimensional dilaton gravity theory. 
The Lagrangian density of the scalar K-field is carefully chosen such that the field equations take the same form as those of the canonical model, and the Schr\"odinger-like equation for the linear perturbation can be obtained without doing coordinate transformation. We found that when the background solution is a sine-Gordon kink, the linear perturbation equation becomes a Schr\"odinger-like equation with the exactly solvable PT II potentials $V_{\textrm{PT},+}(x;l,1)$, where the parameter $l$ depends on the gravitational coupling and the expectation value of the vacuum.

This potential is singular at $x=0$, which leads to some exotic properties. For example, although the Hamiltonians correspond to $V_{\textrm{PT},+}(x;l,1)$ and $V_{\textrm{PT},-}(x;l,1)$ are both factorizable, there is always a bound state of $V_{\textrm{PT},-}(x;l,1)$ with negative eigenvalues $E_{-,0}^{(l,1)}=1 - 2 l$. The spectra of $V_{\textrm{PT},+}(x;l,1)$ and $V_{\textrm{PT},-}(x;l,1)$ are largely different, only parts of their states are degenerated. For  asymmetric singular potentials, such as the Morse potentials, the spectra of the partner potentials will be completely different~\cite{PanigrahiSukhatme1993}. 

The bound state spectrum of $V_{\textrm{PT},+}(x;l,1)$ is $(l-1)^2-(l-2n-3)^2$, where  $n=0,1,2,\cdots<\frac{l-3}{2}$. Thus, when $l>3$, $V_{\textrm{PT},+}(x;l,1)$ begins to capture bound states. Each of these bound states corresponds to a vibrational mode around the kink. We considered an explicit case with $l=4$, for which $V_{\textrm{PT},+}(x;4,1)$ has one bound state, whose wave function is vanished at $x=0$, as forced by the strong singularity of the potential. More bound states will appear as $l$ increases.

This exactly solvable model might be useful for studying the quantization and the collisions of gravitating kinks. The method of the present work can also be extended to other higher-dimensional scalar-gravity models. For example, one may try to constructing cosmic inflation or brane world models with exactly solvable linear perturbation equations by introducing K-fields. It is worth to mention that the perturbation issue of the present work corresponds to the scalar perturbation problems of higher-dimensional models.  With these higher-dimensional models, one may ask what are the possible phenomenological implications of the hard singularity as well as the vibrational modes. We leave these issues for our future works.

\section*{Acknowledgments}
I would like to thank Jun Feng, Bin Guo, Kei-ichi Maeda, Jian Wang and Pengming Zhang for stimulating discussions. This work was supported by the National Natural Science Foundation of China (Grant Number: 12175169), Fundamental Research Funds for the Central Universities (Grant Number:~xzy012019052).

\providecommand{\href}[2]{#2}\begingroup\raggedright\endgroup

\end{document}